\documentstyle[preprint,prc,aps,psfig]{revtex}
\begin{document}
\draft
\title{Elliptic flow from a parton cascade}
\author{Bin Zhang$^{a}$, Miklos Gyulassy$^{b}$, and Che Ming Ko$^{a}$}
\address{$^{a}$Cyclotron Institute and Physics Department,\\
Texas A\&M University, College Station, TX 77843-3366, USA\\
$^{b}$ Physics Department, Columbia University,\\
New York, NY 10027, USA}
\date{February 5, 1999}
\maketitle

\begin{abstract}
\noindent The dependence of elliptic flow at RHIC energies
on the effective parton scattering cross section is calculated
using the ZPC parton cascade model. We show that the $v_2$ measure
of elliptic flow saturates early in the evolution before the
hadronization transition to a rather large value $\sim 0.05-0.15$
as $\sigma_{g}$ varies from 2-10 mb and thus is a sensitive probe
of the dynamics in the plasma phase.
\end{abstract}
\vspace{0.5cm}

\narrowtext


In nucleus-nucleus collisions at the Relativistic Heavy Ion
collider (RHIC), a system of deconfined quarks and gluons is
expected to be produced for about $3-10$ fm/c\cite{mgyul1}. Various
signatures have been proposed to verify its existence
\cite{harris}. Since the volume and lifetime of this matter may be
much larger and longer than those given by the confinement scale
$\approx 1/\Lambda_{QCD}$, it is expected that collective motion of
these deconfined partons should arise and have observable
consequences that provide information on the dynamics in quark
gluon plasma.

The collective motion of particles in heavy ion collisions can be
studied via directed and elliptic flows. The directed flow, which
measures the collective motion of particles in the reaction plane,
has been studied extensively at BEVALAC, SIS, and AGS energies and
found to be appreciable \cite{drisc1,mgyul2,gqli,bali1}. It becomes
small at higher energies due to the large beam rapidity. In this
case, it is more suitable to study the elliptic flow
\cite{sorge1,jolli1}, which measures the azimuthal asymmetry of the
transverse flow pattern. For heavy ion collisions with fixed
targets, such as at CERN SPS, elliptic flow is generated by
hadronic final state interactions \cite{jolli1}. On the other hand,
at RHIC energies and beyond, copious mini-jet production leads to a
high density of partons on a very short time scale and elliptic
flow can be generated from interactions at the partonic level prior
to hadronization. The aim of this letter is to calculate this
effect using the parton cascade code ZPC \cite{bzhan1,bzhan2}.

Initial partonic conditions can be estimated for Au-Au collisions
at RHIC energies via the HIJING generator\cite{hijing1} for
example. Here we will consider simple idealized geometries for a
first analysis. Initially, partons are uniformly distributed in the
overlapping region of two disks each having a transverse radius of
$5$ fm with their centers separated by a distance equal to the
impact parameter. For collisions of large nuclei at not very large
impact parameters, the multiplicity is roughly proportional to the
overlapping area. HIJING estimates for $b=0$ a gluon rapidity
density ${\rm d}N_g/{\rm d}y \approx 300$. Again as a first
idealization we assume a uniform rapidity distribution from $-5$ to
$+5$ and take a momentum distribution given by a local thermal one
at a temperature of $500$ MeV. Furthermore, produced particles are
taken to be formed on a hyperbola in the $t$-$z$ plane, with $z$
the beam direction and $x$-$z$ plane the reaction plane, using a
formation proper time $\tau_0=0.2$ fm.

At RHIC energies, the initial minijet system is dominated by
gluons. Their subsequent time evolution can be described by the
parton cascade model ZPC. In this model, only gluon-gluon elastic
scattering has been included, with a cross section
\[\frac{{\rm d}\sigma_{g}}{{\rm d}t}
=\frac{9\pi\alpha^2}{2}\left(1+\frac{\mu^2}{s}\right)
\frac{1}{(t-\mu^2)^2}.\]
In the above, the strong interaction coupling constant $\alpha$ is
taken to be 0.47, and $\mu$ is an effective screening mass
responsible for regulating the divergent leading-order cross
section. We vary $\mu$ to study the dependence of the elliptic flow
on $\sigma_{g}$.

The elliptic flow of partons is characterized by the second Fourier
coefficient $v_2$ in their momentum distribution, i.e.,
\begin{equation}
\frac{1}{N}\frac{{\rm d}N}{{\rm d}\phi}
= v_0 + 2 v_1 \cos(\phi) + 2 v_2 \cos(2 \phi) +
\cdots.
\end{equation}
Alternatively, it can be characterized by $v_2'$ in the expansion
of their transverse energy distribution,
\begin{equation}
\frac{1}{\langle E_T\rangle}\frac{{\rm d}E_T}{{\rm d}\phi}
= v_0' + 2 v_1' \cos(\phi) + 2 v_2' \cos(2 \phi) +
\cdots.
\end{equation}
Numerically, $v_2$ is measured by averaging
\[\cos(2\phi) = \frac{p_x^2-p_y^2}{p_x^2+p_y^2}\]
over partons, i.e., it is the single particle average of squared
transverse momentum asymmetry. Similarly, $v_2'$ is measured by the
$E_T$ weighted single particle average of $\cos(2\phi)$. In
general, the two azimuthal distributions do not have to be the
same, and the two coefficients $v_2$ and $v_2'$ are thus not
necessarily equal \cite{wreis1}. We will later give an example in
which they develop different values.

The elliptic flow arises due to multiple collisions as long as the
initial transverse spatial distribution is azimuthally asymmetric.
For fixed impact parameters, its value reflects the strength of
interactions ($\sigma_g$). For fixed $\sigma_g$ its value depends
on the spatial asymmetry as a function of the impact parameter (or
multiplicity).

In Fig. 1, we first show results for fixed impact parameter $b=7.5$
fm. The time evolution of $v_2$ for different values of the
gluon-gluon scattering cross section is shown. Note that all
asymptotic values are reached very early, $t<2$ fm/c, well before
the hadronization transition. For interaction length
$\sqrt{\sigma/\pi}$ larger than the mean free path, the numerical
method of particle partition \cite{chart1} has been used in order
to obtain stable solutions of the Boltzmann equation. For the
typical cross section between gluons, 3 mb, the final $v_2$ after
the partonic stage is around $0.1$ which is quite ``large" and
easily detectable. Since the screening mass, or equivalently, the
effective parton scattering cross section is uncertain we study its
effect on the elliptical flow by using different values for the
cross section, i.e., 1 mb and 10 mb. We see that $v_2$ is rather
sensitive to the cross section. This demonstrates that very large
dissipative corrections are at work since in the ideal fluid limit
the flow depends only on the speed of sound that is in our case
$1/\sqrt{3}$. The important role played by dissipative phenomena in
collective properties of plasma evolution was also demonstrated in
\cite{bzhan2} in connection with the transverse energy evolution.
For these initial conditions even Navier-Stokes is inadequate and
the full microscopic parton cascade is necessary. The strong
dependence of the elliptic flow on the parton cross section could
provide an important probe of the plasma dynamics if hadronization
and final hadronic rescatterings do not modify the results too
much. We cannot yet address this problem within the scope of ZPC
and leave that important topic to a future study.

In Fig. 2, we compare the time evolution of $v_2$ and $v_2'$ for a
gluon-gluon cross section of 10 mb. The statistical errors are
denoted by bars. We see that $v_2'$ is always distinctively larger
than $v_2$ during the evolution. It indicates that particles moving
in the $x$ direction tend to have relatively larger $E_T$ than
particles moving in the $y$ direction. Even though final-state
particle production may change the difference, this gives an
example of different values for $v_2$ and $v_2'$. The difference
between $v_2$ and $v_2'$ becomes less significant when the cross
section becomes smaller.

To study how the elliptic flow reflects the collision geometry, we
show in Fig. 3 the dependence of the elliptic flow on the impact
parameter for medium range impact parameters. An almost linear
dependence is observed, and this is similar to the impact parameter
dependence obtained from the hydrodynamic model \cite{jolli1}. The
linear increase of elliptic flow as a function of impact parameter
reflects the initial spatial asymmetry, which can be characterized,
e.g., by the width $L_x$ and height $L_y$ of the overlapping
region, via $\alpha_s=(L_y-L_x)/(L_y+L_x)$. One can easily show
that for medium range impact parameters, the spatial asymmetry is
approximately linearly proportional to the impact parameter,
similar to the impact parameter dependence of the elliptical flow.
However, for $b/(2R)$ very close to $1$, the spatial asymmetry
approaches 1, but the $v_2$ coefficient actually drops to 0. This
is simply due to the fact that there are too few particles which
can undergo scattering to develop a momentum asymmetry from the
spatial asymmetry. In Fig. 4, the time evolution of $v_2$ is shown
for different impact parameters. Although collisions at larger
impact parameter give rise to a larger $v_2$, it reaches the final
value also sooner, indicating that it is the size of the
overlapping region in the $x$ direction which determines how fast
the elliptical flow is generated in the collision.

To summarize, we have studied in this letter the elliptical flow
generated during the partonic evolution in heavy ion collisions at
RHIC energies. For medium range impact parameters, an appreciable
elliptical flow has been found as a result of the initial spatial
asymmetry and subsequent partonic rescattering. The magnitude of
the elliptical flow shows a strong dependence on the parton-parton
scattering cross section, demonstrating the importance of
dissipative corrections to idealized hydrodynamic expectations. The
finite parton elliptical flow is expected to lead to an initial
hadron matter with an asymmetric azimuthal momentum distribution.
This is different from heavy ion collisions at lower energy, such
as the SPS, where the hadronic evolution starts with an
approximately isotropic azimuthal momentum distribution in the
nuclear overlapping region. Since the spatial asymmetry is
relatively smaller during the hadronic evolution in heavy ion
collisions at RHIC, one does not expect that hadronic rescattering
will generate appreciable elliptical flow \cite{sorge1} and thus
should not significantly modify the initial partonic elliptical
flow.

In this first study we have not yet investigated the effects due to
different initial conditions for the partonic matter. For example,
the turbulent glue initial conditions given by the HIJING model
\cite{mgyul3} may lead to a different value for the elliptical flow
as obtained here using a thermal one. A quantitative study of
effects due to both the hadronic rescattering and the initial
parton distribution is underway by using a transport model that
takes results from the HIJING model \cite{hijing1} as input to the
ZPC parton cascade model \cite{bzhan1} and then includes the hadron
cascade via the ART model \cite{bali1}. Such a comprehensive study
will allow us to find quantitatively the sensitivity of elliptical
flow on the parton-parton scattering cross section, thus making it
possible to extract valuable information on the properties of the
hot dense parton matter formed in the initial stage of
ultrarelativistic heavy ion collisions.

We thank B.A. Li for helpful discussions. The work of BZ and CMK
was supported in part by the National Science Foundation under
Grant No. PHY-9870038, the Welch Foundation under Grant No. A-1358,
and the Texas Advanced Research Program. The work of MG was
supported by the U.S. Department of Energy under contract No.
DE-FG02-93ER40764.

{}

\clearpage

\listoffigures

\clearpage

\begin{figure}[h]
\vspace{5cm}
\hspace{0.5cm}
\psfig{figure=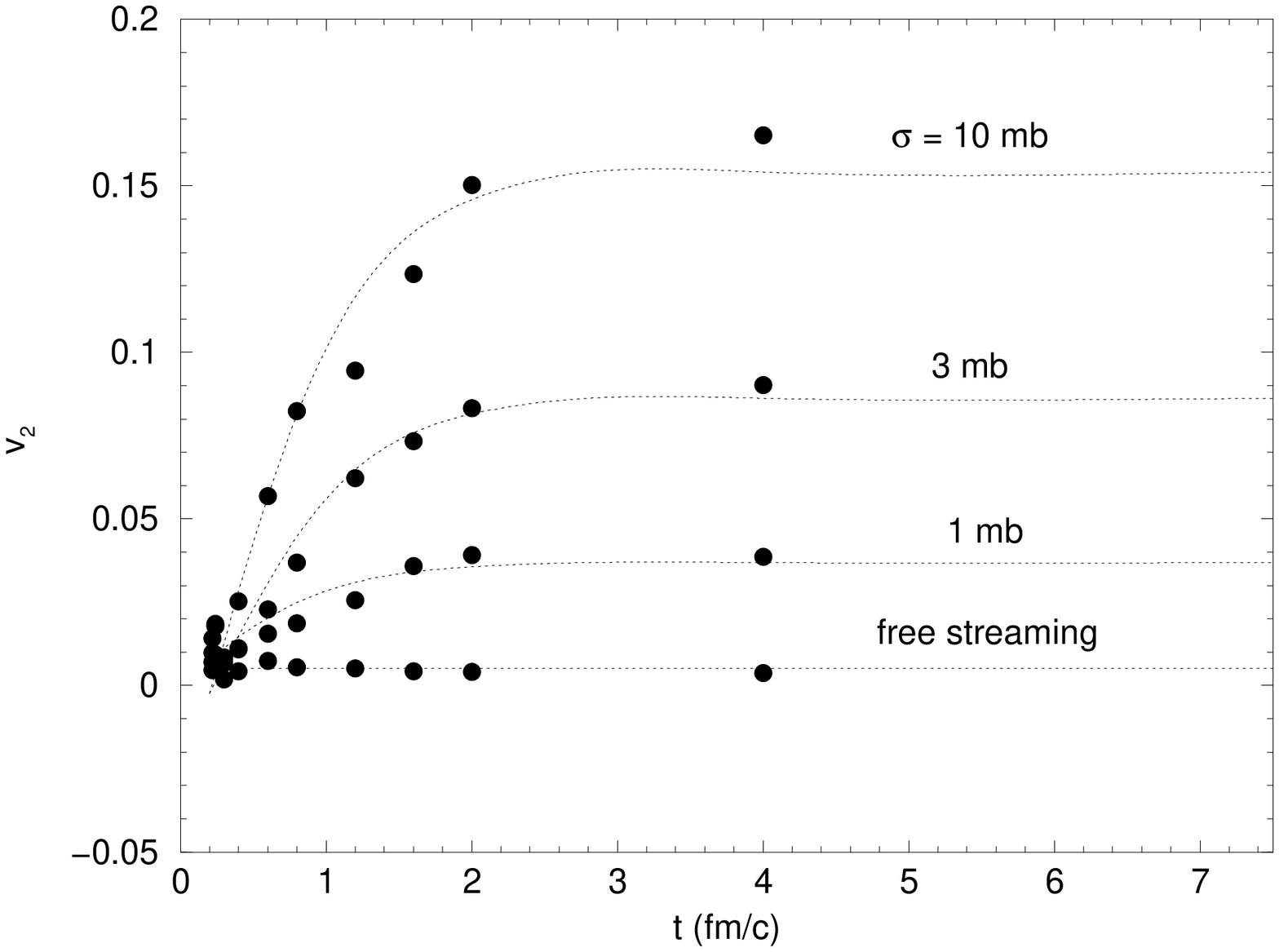,height=3.5in,width=3.0in,angle=-90}
\caption{
Time evolution of $v_2$ coefficient for different effective parton
scattering cross sections in Au-Au collisions at
$\protect{\sqrt{s}}=200$ AGeV with impact parameter $7.5$ fm.
Filled circles are cascade data, and dotted lines are hyperbolic
tangent fits to the data. }
\end{figure}

\clearpage

\vspace{2cm}
\hspace{0.5cm}
\begin{figure}[h]
\vspace{5cm}
\hspace{0.5cm}
\psfig{figure=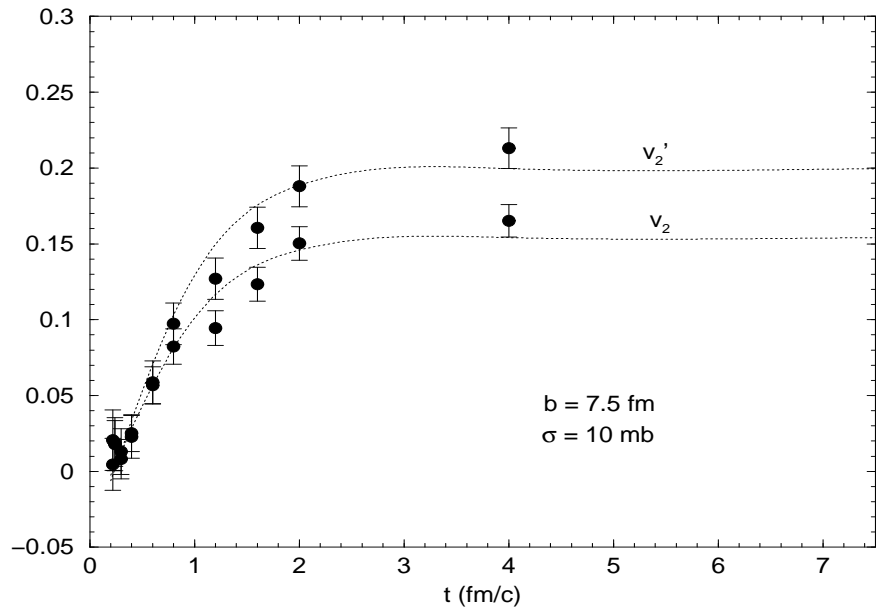,height=3.5in,width=3.0in,angle=-90}
\caption{
Time evolution of $v_2$ and $v_2'$ coefficients from the same
reaction as in Fig. 1 for $\sigma=10$ mb and $b=7.5$ fm. Filled
circles are cascade data with statistical errors, and dotted lines
are hyperbolic tangent fits to the data. }
\end{figure}
\clearpage

\vspace{2cm}
\hspace{0.5cm}
\begin{figure}[h]
\vspace{5cm}
\hspace{0.5cm}
\psfig{figure=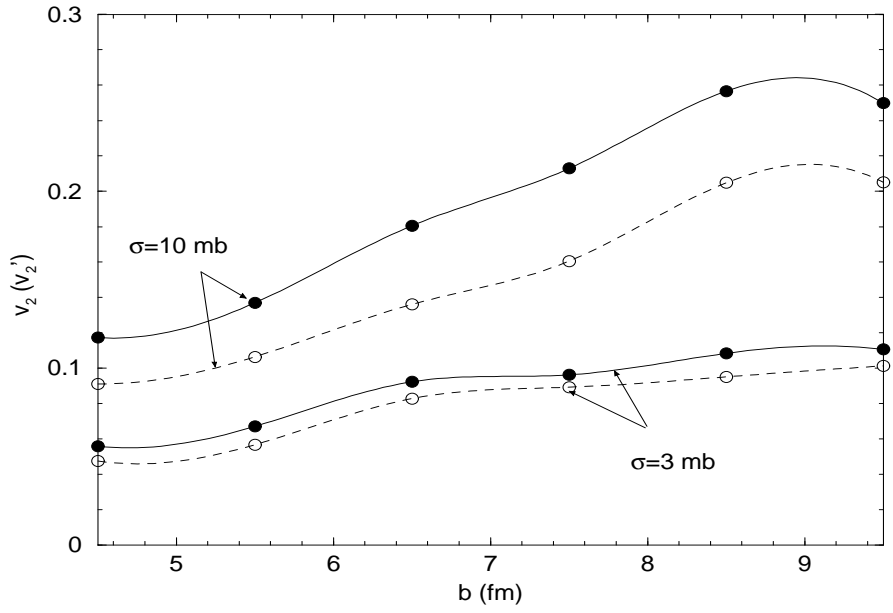,height=3.5in,width=3.0in,angle=-90}
\caption{
Impact parameter dependence of $v_2$ and $v_2'$. Filled circles are
for $v_2'$ and open circles are for $v_2$. The lines are splines to
guide the eyes. }
\end{figure}

\clearpage

\vspace{2cm}
\hspace{0.5cm}
\begin{figure}[h]
\vspace{5cm}
\hspace{0.5cm}
\psfig{figure=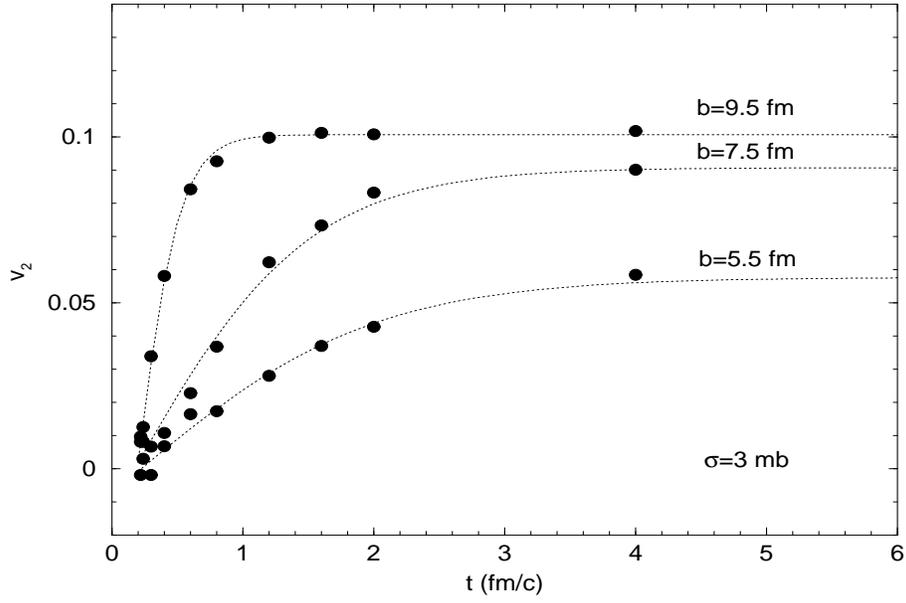,height=3.5in,width=3.0in,angle=-90}
\caption{
Time evolution of $v_2$ for different impact parameters. The cross
section is taken to be $3$ mb. Filled circles are cascade data, and
dotted lines are hyperbolic tangent fits to the data. }
\end{figure}

\end{document}